\documentclass[conference]{IEEEtran}
\usepackage[T1]{fontenc}
\IEEEoverridecommandlockouts
\usepackage{cite}
\usepackage{amsmath,amssymb,amsfonts}
\usepackage[linesnumbered,ruled,vlined]{algorithm2e}
\usepackage{graphicx}
\usepackage{textcomp}
\usepackage{algpseudocode}
\usepackage{xcolor}
\usepackage{booktabs}
\usepackage{tabularx}
\def\BibTeX{{\rm B\kern-.05em{\sc i\kern-.025em b}\kern-.08em
    T\kern-.1667em\lower.7ex\hbox{E}\kern-.125emX}}

\definecolor{matt}{RGB}{219, 48, 122}
\definecolor{jad}{RGB}{0,0,255}

\begin{document}

\title{Under Pressure: Security Analysis and Process Impacts of a Commercial Smart Air Compressor}

\author{\IEEEauthorblockN{Jad Zarzour}
\IEEEauthorblockA{
\textit{Mason Innovation Labs}\\
\textit{George Mason University}\\
Arlington, Virginia\\
jzarzour@gmu.edu}
\and
\IEEEauthorblockN{Matthew Jablonski}
\IEEEauthorblockA{
\textit{Mason Innovation Labs}\\
\textit{George Mason University}\\
Arlington, Virginia\\
mjablons@gmu.edu}
}

\maketitle

\begin{abstract}
The integration of Industrial Internet of Things (IIoT) devices into manufacturing environments has accelerated the transition to Industry 4.0, but has also introduced new cybersecurity risks. This paper conducts a comprehensive security analysis of a commercial smart air compressor, revealing critical vulnerabilities including hardcoded credentials, unauthenticated APIs, and an insecure update mechanism. It includes a formal threat model, demonstrates practical attack scenarios in a testbed environment, and evaluates their subsequent impact on an industrial process, leading to denial of service and the corruption of critical process telemetry. In addition, an analysis of the device’s supply chain reveals how product integration from multiple vendors and limited security considerations can expose a device to threats. The findings underscore the necessity of incorporating cybersecurity principles into both IIoT device design and supply chain governance to enhance resilience against emerging industrial cyber threats.

 \end{abstract}
\section{Introduction}
\label{sec:intro}
Industrial development is increasingly defined by the principles of Industry 4.0, which integrates innovations such as big data analytics, simulation, internet services, cyber-physical systems (CPS), and the Internet of Things (IoT). 
Despite the significant advancements these technologies promise, their secure implementation in factory and industrial environments face numerous challenges.

The adoption of IoT devices has expanded in both consumer and industrial sectors.
In industrial applications, the Industrial Internet of Things (IIoT) involves deploying connected sensors and actuators that are critical to factory operations.
This paper focuses on smart air compressors, a foundational utility that powers pneumatic tools, automated systems, and material handling.
Market analyses indicate that manufacturing represents the largest end-use segment for air compressors, accounting for a 34\% revenue share in 2024 and highlighting their operational importance~\cite{MordorIntel2025}. 

The convergence of information technology (IT) and operational technology (OT) has created significant and costly cybersecurity risks. 
According to IBM's 2025 Cost of a Data Breach Report, breaches affecting IIoT and OT environments add an average of \$175,010 to the total cost of a incident.
Additionally, supply chain compromise has emerged as the second most prevalent attack vector, accounting for 15\% of all breaches and increasing costs by an average of \$227,224~\cite{IBM2025}.
Embedded IIoT devices are of particular concern, as they often lack adequate security controls despite their critical roles~\cite{SisinniEmiliano2018IIoT}.


This paper makes the following three contributions. First, it presents the first comprehensive security analysis of a commercial smart air compressor, demonstrating practical attacks that compromise pneumatic system availability and integrity. Second, it provides a formal threat model grounded in the ISA/IEC 62443 and NIST SP 800-82 Rev. 3 frameworks, offering a reusable methodology for evaluating similar IIoT devices. Lastly, it analyzes supply chain security risks through a detailed case study and offers concrete and actionable recommendations for manufacturers and system integrators.

This paper is structured as follows: In Section~\ref{sec:background}, we provide a background on compressed air systems in industrial environments. In Section~\ref{sec:systhreatmodel}, we present our formal threat model for analyzing the security of a target smart air compressor. Section~\ref{sec:experiment} describes our experimental methodology, and Section~\ref{sec:evaluation} provides the results of experimentation. In Section~\ref{sec:defense}, we provide defensive mitigation recommendations in developing smart air compressors. Section~\ref{sec:supplychain} describes our supply chain case study. Section~\ref{sec:relatedwork} describes related work and Section~\ref{sec:conclusion} provides conclusions.

\section{Background}
\label{sec:background}
Compressed air is often called the \textit{fourth utility} in manufacturing~\cite{Koski01052002}, along with electricity, water, and natural gas, underscoring its critical role in production.
Factories rely on air compressors for various industrial purposes, including refrigeration, pneumatic transport, and power production~\cite{tamilselvi2025compressor}. 
Because compressed air drives safety-critical processes, any disruption or tampering carries severe risks, from manufacturing line shutdowns and pneumatic brake failures to a loss of precise pressure control that could damage equipment and endanger workers~\cite{zunjic2024analysis}.

Given these operational and safety risks, governments have implemented regulations for compressed air systems. 
In the United States, the Occupational Safety and Health Administration (OSHA) mandates safety features to mitigate hazards at all levels of compressed air systems. For example, 29 CFR 1910.169(b)(3) requires that air tanks have drain valves to prevent the accumulation of excessive liquid in the receiver~\cite{OSHA1910_169b2}. 
Further, 29 CFR 1910.430 (b) requires that all air compressor systems have a check valve on the inlet side~\cite{OSHA1910_430b} to allow airflow into the compressor and prevent backflow when the compressor is off or when the system pressure exceeds the suction side pressure.

Air compressors have begun to be integrated into smart factories as IoT device, they integrate sensors and data analytics for real-time monitoring. By transmitting this data to industrial networks, operators can gain actionable insights into system performance, automate control functions, and ensure the reliability of critical pneumatic operations across manufacturing, energy, and infrastructure sectors.

From a cybersecurity perspective, these physically critical systems represent significant cyber-physical targets. 
Modern air compressors increasingly include IIoT devices for health monitoring and optimization as the industry seeks to reduce costs~\cite{murty2019compressor}.
However, integrating wireless connectivity into traditionally isolated industrial equipment introduces security challenges unique to the OT environment.
Unlike enterprise IT systems, IIoT devices often have limited resources and lifespans exceeding a decade, making traditional security difficult to maintain.
While prior security assessments of similar devices exist~\cite{pricop2014assessing}, they provide only a surface-level analysis and do not investigatie practical attack methodologies that exploit common weaknesses like weak authentication, unencrypted communications, and insufficient input validation.
\section{System and Threat Model}
\label{sec:systhreatmodel}

This section describes the smart air compressor system in an industrial context and defines a formal threat model for its analysis.

\subsection{Target System Overview}
\label{ssec:target}

\begin{figure}
    \centering
    \includegraphics[width=1\linewidth]{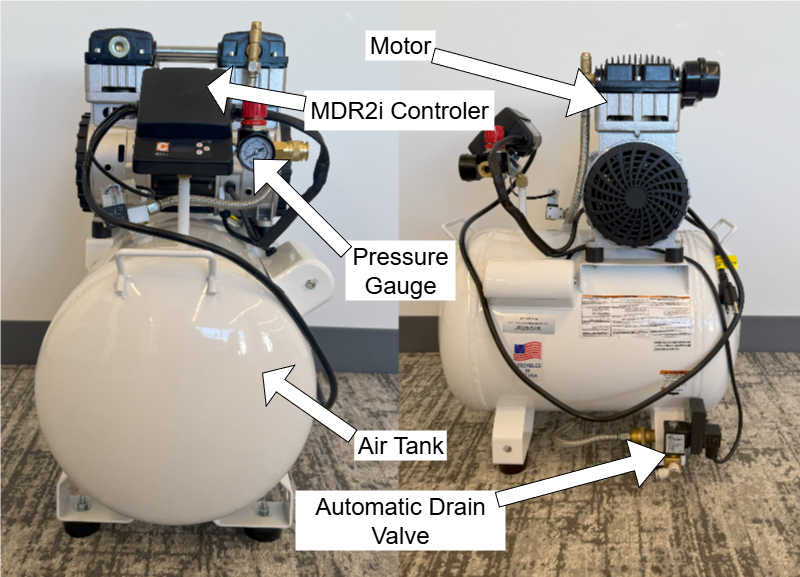}
    \caption{California Air Tools CAT-10020SMHAD smart air compressor with MDR2i wireless controller.}
    \label{fig:aircompressor}
\end{figure}

The target platform is the California Air Tools CAT-10020SMHAD, equipped with the MDR2i smart controller that provides remote monitoring and Wi-Fi control (Figure~\ref{fig:aircompressor})~\cite{CATpage2025}. 
In this study, the compressor supplies air to a workcell controlled by a Programmable Logic Controller (PLC).
The MDR2i exposes its configuration and status via a web interface and supports over-the-air (OTA) firmware updates through its user interface (UI), creating a trust boundary relevant to this model~\cite{CATmdr2i2025}. 
The device documentation specifies configurable cut-in/cut-out thresholds with a default working pressure of 90-120 PSI, adjustable up to 105 PSI kick-on and 130 PSI kick-off~\cite{CATmanual2025}.

\subsection{Assets and Interfaces}

Primary assets for this analysis include the MDR2i controller firmware, its configuration store, the web application, the Wi-Fi management plane, the start/stop control path, and telemetry data (pressure, current, run-time counters) exposed via the web UI. 
External interfaces include the on-device OLED display and controls, the wireless network, web endpoint, and OTA update mechanism accessible from the controller's UI. 
Each interface represents a trust boundary distinct from the compressor's electro-pneumatic domain.

From a cybersecurity assurance perspective, the MDR2i controller aligns with the \textit{embedded device} component type in the ISA/IEC 62443 Part 4-2 standard~\cite{IEC62443-4-2-2019}.
This standard specifies security requirements for industrial automation and control system (IACS) components, organized into seven Foundational Requirements (FRs).
For an embedded device such as the MDR2i, these requirements include capabilities for authentication, update integrity, and configuration management. 
For example, FR1 (identification and authentication control) and FR3 (system integrity) specify component-level controls to authenticate users and devices, verify software and firmware authenticity, and ensure the integrity of updates~\cite{ISCI2024}. 
These examples are directly relevant to the vulnerabilities analyzed in this paper.

\subsection{Threat Model}
\label{ssec:threatmodel}

This model defines the scope, attacker capability, assets, security properties, trust boundaries, objectives, and attack vectors that are evaluated in the subsequent sections.

\begin{figure*}
    \centering
    \includegraphics[width=1\linewidth]{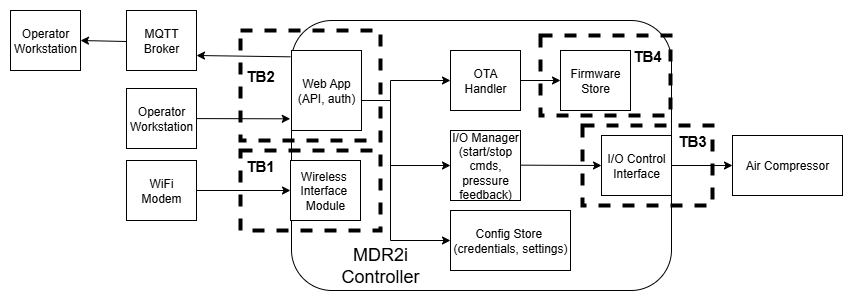}
    \caption{Smart air compressor system architecture showing trust boundaries $TB_1$ (wireless interface vs. internal controller logic), $TB_2$ (web application vs. configuration store), $TB_3$ (controller electronics vs. electro-pneumatic domain), and $TB_4$ (OTA update path vs. firmware integrity verification). The MDR2i controller contains six key modules that handle wireless communication, web services, configuration management, firmware updates, and pneumatic control interfaces.}
    \label{fig:trustbound}
\end{figure*}

\paragraph{Scope and Assumptions} The analysis focuses on network-based attacks and excludes physical manipulation, destructive overpressure testing, or attacks requiring specialized ICS protocol knowledge.
We assume a standard workcell configuration where pneumatic systems operate critical actuators and pressure telemetry feeds into supervisory monitoring systems.

\paragraph{Threat Actors and Capabilities} The adversary has local network access to the controller, within Wi-Fi AP range or on the same LAN segment in station mode, without prior credentials or physical access. The baseline profile corresponds to ISA/IEC 62443 Security Level 1 (SL-1), meaning protection against casual or coincidental violation by an actor with simple means, few resources, generic skills, and low motivation. All attacks evaluated are feasible under these SL-1 assumptions.

\paragraph{Assets} Critical assets are defined as: ($AS_1$) pneumatic air availability to the workcell, ($AS_2$) integrity of pressure telemetry data used by the digital twin, ($AS_3$) confidentiality of system configuration and credentials, and ($AS_4$) authorized start/stop and setpoint control over compressor operation.

\paragraph{Security Properties} Properties to be preserved are ($SP_1$) availability of continuous air supply, ($SP_2$) integrity of accurate and timely pressure readings, and ($SP_3$) authenticity and confidentiality of control commands and system credentials.


\paragraph{Trust Boundaries} Following the guidance from NIST SP 800-82 Rev. 3 on OT network segmentation and boundary protection~\cite{NIST-SP800-82r3}, and using the guidance from ISA/IEC 62443 on zones and conduits, we treat the following interfaces as distinct trust boundaries: ($TB_1$) the wireless interface vs. internal controller logic, ($TB_2$) the web application vs. its configuration/data store, ($TB_3$) the controller electronics vs. the electro-pneumatic domain (field I/O), and ($TB_4$) the OTA update path vs. firmware integrity verification.
These trust boundaries are depicted in Figure~\ref{fig:trustbound}.

\paragraph{Attacker Goals} Objectives are categorized using MITRE ATT\&CK for ICS: ($O_1$) \textit{Impact} - disrupt availability or control of pneumatic supply to cause downtime, ($O_2$) \textit{Impair Process Control} - manipulate pressure telemetry to mislead monitoring, and ($O_3$) \textit{Persistence} - establish continued access for repeated disruption~\cite{alexander2020attackics}.

\paragraph{Attack Vectors} The evaluated attack vectors are derived directly from the trust boundaries to achieve the attacker's goals assuming SL-1 capabilities within the established scope. 
Let $AS$, $SP$, $TB$, and $O$ respectively be the labeled sets of assets, security properties, trust boundaries, and attacker objectives, and define a single relation $V \subseteq TB \times SP \times AS \times O$  where $(tb, sp, as, o) \in V$ indicates that violating security property $sp$ for asset $as$ through boundary $tb$ achieves objective $o$ under the stated SL-1 assumptions.

\begin{table}
    \caption{The threat vector relation $V$ induced by SL-1 attacks.}
    \label{tab:vectors}
    \centering
    \begin{tabular}{lllll}
    \toprule
    \textbf{TB} & \textbf{SP} & \textbf{AS} & \textbf{O} & \textbf{Vector Description} \\
    \midrule
    $TB_1$ & $SP_1$ & $AS_1$ & $O_1$ & Wireless misconfig. \\
    $TB_2$ & $SP_1$ & $AS_1$ & $O_1$ & Reset/restart abuse \\
    $TB_2$ & $SP_2$ & $AS_2$ & $O_2$ & Unauth. calibration \\
    $TB_2$ & $SP_3$ & $AS_4$ & $O_1$ & Unauth./weak creds. (control) \\
    $TB_2$ & $SP_3$ & $AS_4$ & $O_3$ & Unauth./weak creds. (persist.) \\
    $TB_3$ & $SP_1$ & $AS_1$ & $O_1$ & Unsafe controller–plant \\
    $TB_4$ & $SP_3$ & $AS_4$ & $O_1$ & OTA manipulation (impact) \\
    $TB_4$ & $SP_3$ & $AS_4$ & $O_3$ & OTA manipulation (persist.) \\
    \bottomrule
    \end{tabular}
\end{table}

The relation $V$ induced by the vectors in this study is detailed in Table~\ref{tab:vectors}.
This formal approach enables systematic evaluation and comparison with other IIoT security assessments.
The threat vector relation $V$ provides a structured framework for prioritizing defensive measures based on trust boundary violations, supporting quantitative risk assessment practices recommended by NIST SP 800-82 Rev. 3~\cite{NIST-SP800-82r3}. 
While Table~\ref{tab:vectors} provides a formal, high-level mapping of these threat vectors, a comprehensive traceability matrix is presented after the technical analysis in Table~\ref{tab:traceability} to connect these abstract vectors to the concrete attacks and defenses detailed in this work.


\begin{figure*}
    \centering
    \includegraphics[width=\linewidth]{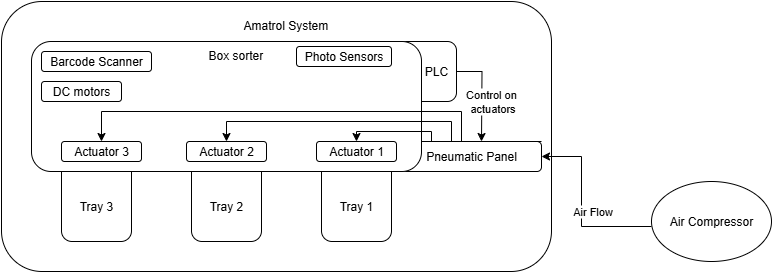}
    \caption{Amatrol pneumatic sorting workcell used in the experimental design, showing the PLC-controlled barcode scanner, photo sensors, conveyor motors, three pneumatic actuator feeding trays, and the smart air compressor supplying ~100 PSI.}
    \label{fig:amatrol}
\end{figure*}

\section{Experimental Methodology}
\label{sec:experiment}

This section details the laboratory testbed configuration and the attacker environment used to evaluate the security of the CAT-10020SMHAD smart air compressor under realistic industrial conditions.

\subsection{Testbed and Instrumentation}
\label{ssec:testbed}

The experimental testbed simulates a representative industrial workcell environment, as shown in Figure~\ref{fig:amatrol}.
The testbed consists of a PLC-controlled conveyor belt system designed to sort packages based on barcode labels. 
Pneumatic actuators, requiring a compressed air supply of ~100 PSI, push packages from the conveyor to one of three designated trays. 
This setup allows for an end-to-end evaluation of how attacks targeting the smart air compressor affect a representative industry process.

\begin{figure}
    \centering
    \includegraphics[width=1\linewidth]{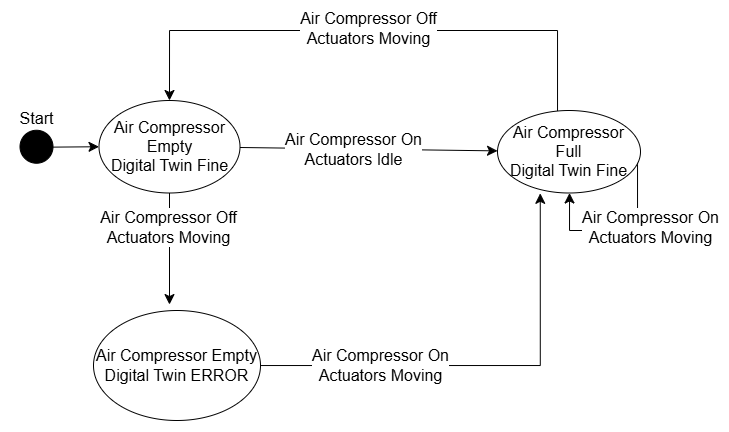}
    \caption{Machine State Diagram of the system with a physical air compressor. Without pressure telemetry, the digital twin unreliably infers package movement from actuator counters alone.}
    \label{fig:StateDiagramNoIoT}
\end{figure}

\begin{figure}
    \centering
    \includegraphics[width=1\linewidth]{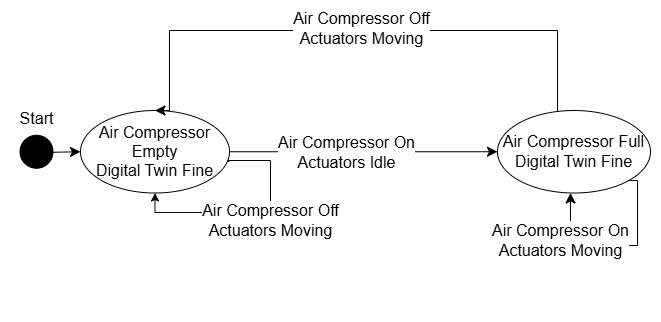}
    \caption{Machine State Diagram of the system with the integrated SMART air compressor. Real-time pressure telemetry allows the digital twin to correlate actuator commands with actual pneumatic pressure, improving state awareness.}
    \label{fig:StateDiagramWithIoT}
\end{figure}

The system's digital twin, which monitors package locations via PLC data, was enhanced for this study.
Initially, the digital twin could only infer package movement by monitoring actuator counters, a method that proved unreliable if the air supply was unavailable, as modeled in the state diagram in Figure~\ref{fig:StateDiagramNoIoT}.
To address this limitation, we integrated the CAT-10020SMHAD smart air compressor, connecting its MDR2i controller to the digital twin via its \texttt{/parameter} API endpoint. 
This integration provides real-time pressure telemetry, allowing the digital twin to correlate actuator commands with actual pneumatic pressure, improving state awareness as shown in Figure~\ref{fig:StateDiagramWithIoT}.   

The testbed is isolated on a dedicated VLAN, and a separate SSID was used for testing the smart air compressor. 
The MD2Ri controller firmware under test was version 2501281017, dated January 28, 2025.

\subsection{Attack Implementation}
\label{ssec:attackimplement}

To evaluate the security of the smart air compressor, we established an attack environment using a laptop running Kali Linux, equipped with an ALFA AWUS036AXML Wi-Fi adapter to enable wireless monitoring and packet injection.
We utilized a standard penetration testing toolkit, including: 

\begin{itemize}
    \item \textbf{Kismet and Wireshark} for wireless network discovery, traffic analysis, and full packet capture logging.
    \item \textbf{Burp Suite Professional} for intercepting, analyzing, and manipulating HTTP requests to the controller's web API.
    \item \textbf{Custom Python scripts} for automating brute-force and denial-of-service attacks.
\end{itemize}

This setup aligns with the capabilities of the SL-1 attacker defined in our threat model.


\section{Evaluation}
\label{sec:evaluation}

\begin{figure}
    \centering
    \includegraphics[width=1\linewidth, height=10cm, keepaspectratio]{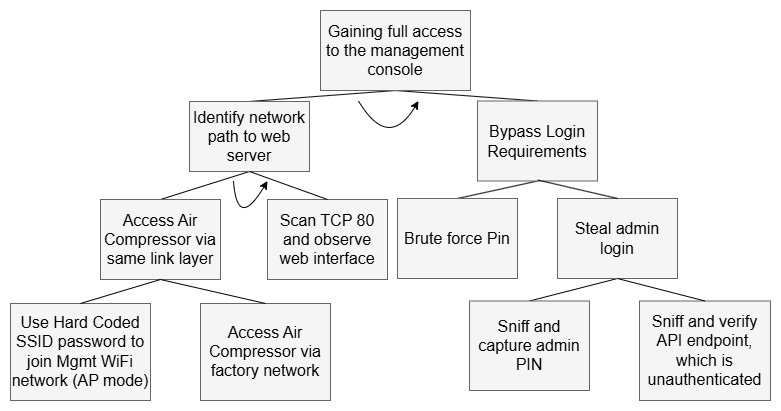}
    \caption{Persistence $O_3$ Attack Tree: Issues with Wi-Fi authentication, coupled with minimal web authentication configurations, allow an SL-1 attacker to use their capabilities to gain admin access to the smart air compressor.}
    \label{fig:persisto3}
\end{figure}

In this section, we evaluate the security of the CAT-10020SMHAD smart air compressor. 
We first describe the methods an SL-1 attacker can use to gain privileged access and then detail practical attacks that impact availability and integrity as shown in Figure~\ref{fig:persisto3}. 

\subsection{Basic Network Access}
\label{ssec:basicnetwork}

($AT_1$) \textbf{Hardcoded Network Credentials:} The MDR2i controller supports two Wi-Fi modes, access point (AP) and station. In its default AP mode, the controller broadcasts an SSID in the format \texttt{MDR2i$\_$<serial-number>}. The corresponding WPA2 password, (\texttt{CATMDR2i}), is published in the user manual and cannot be changed. 
An attacker connecting to the device's network (192.168.50.1) is granted an unauthenticated operator session. 
The hard-coded and unchangeable credential represents a critical vulnerability. 

($AT_2$) \textbf{Station Mode Reconnaissance:} In station mode, the controller connects to an existing Wi-Fi network. An attacker on the same LAN can discover the controller's DHCP-assigned IP address through a simple network scan.


\subsection{Insecure Operator Console}

($AT_3$) \textbf{Unencrypted Control Plane:} After gaining network access, an attacker can identify the operator web console, which serves as the entry point for achieving persistent access (see Figure~\ref{fig:persisto3}). 
A network scan reveals that only port 80 (HTTP) is open. The absence of an HTTPS option means all communications, including credentials and control commands, are transmitted in plaintext. 
This lack of transport-layer security allows for passive eavesdropping by any attacker on the network. 

\subsection{Privilege Escalation}
\label{ssec:privesc}

The web console provides unauthenticated read-only access by default. 
However, we identified a login page for higher-privilege roles. As the user manual did not document these accounts, we contacted the manufacturer's support, who provided a generic account name and a four-digit numerical PIN that appears to be hardcoded across all units. The identified Operator, Manufacturer, and CPC roles and privileges are shown in Table~\ref{tab:roles}.

    \begin{table}[t!]
    \centering
    \caption{MDR2i Controller User Roles and Privileges}
    \label{tab:roles}
    \begin{tabularx}{\columnwidth}{ l X X }
    \toprule
    \textbf{Role} & \textbf{Permissions} & \textbf{Notes} \\
    \midrule
    \textbf{Operator} & 
        \textbullet{} Start/stop the air compressor \newline 
        \textbullet{} Set the target pressure 
        & The default, unauthenticated role. \\
        \addlinespace
    \textbf{Manufacturer} & 
        \textbullet{} Modify the operational pressure range \newline 
        \textbullet{} Configure over/under voltage safety thresholds
        & A privileged role accessed via a shared, hardcoded PIN. \\
        \addlinespace
    \textbf{CPC} & 
        \textbullet{} Change the pressure measurement unit \newline 
        \textbullet{} Recalibrate the pressure sensor zero-point \newline 
        \textbullet{} All lower-level permissions
        & The highest privilege level, also accessed via a shared, hardcoded PIN. \\
    \bottomrule
    \end{tabularx}
    \end{table}


($AT_4$) \textbf{Unrestricted Credential Brute-Force:} To gain CPC-level access, we extracted the credential name from the HTML login form and implemented a basic brute-force script (Algorithm~\ref{alg:bruteforce}) to iterate through all 10,000 possible four-digit PINs. The interface has no rate-limiting or lockout mechanism. 

($AT_5$) \textbf{Use of Shared Static Credentials:} Further, there is no available method to change the PIN for any role. The hardcoded, shared credential scheme is a critical vulnerability. Successful brute-forcing grants an attacker full administrative control over the compressor. 

\begin{algorithm}[t!]
\label{alg:bruteforce}
\caption{Brute-force Password Discovery}
\ForAll{4-digit combinations $p \in \{0000, 0001, \dots, 9999\}$}{
    response $\gets$ POST($URL_{setpass}$, $p$)\;
    \If{response = success}{
        params $\gets$ POST($URL_{getparams}$)\;
        userLevel $\gets$ params[``USERLEVEL'']\;
        \If{$userLevel \neq 0$}{
            Write $(p, userLevel)$ to output file\;
        }
    }
}
\end{algorithm}

\begin{figure}
    \centering
    \includegraphics[width=1\linewidth]{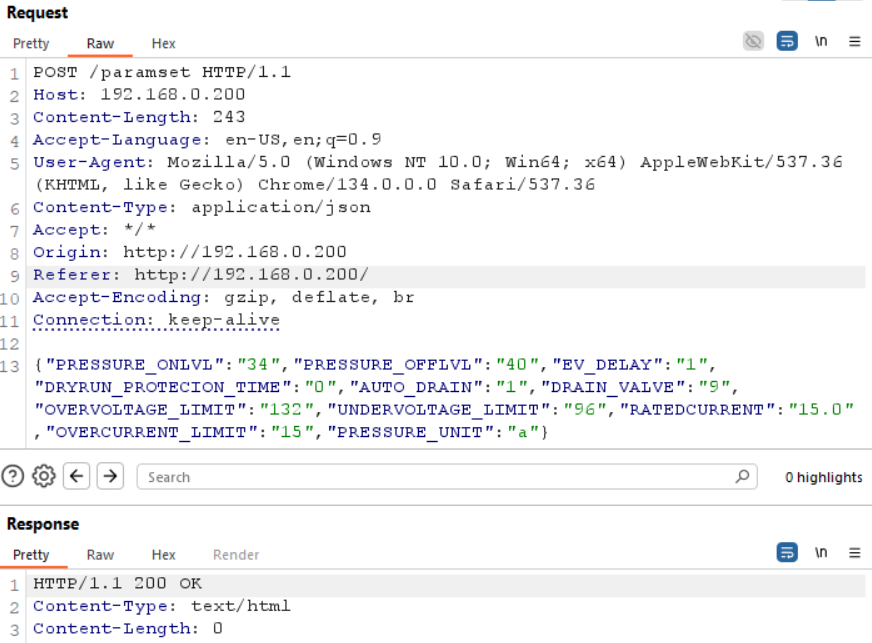}
    \caption{Unauthenticated API access ($AT_6$) demonstrated using Burp Suite. A direct POST request to the \texttt{/setparam} endpoint, sent without any session token or credentials, is accepted with a "200 OK" response, allowing an attacker to manipulate critical safety thresholds. This behavior was consistent across all state-changing API endpoints, including \texttt{/reset} and \texttt{/setpressurerange}.}
    \label{fig:burpsuite}
\end{figure}

\subsection{Unauthenticated APIs}

($AT_6$) \textbf{Missing API Endpoint Authentication:} While the web UI requires a PIN for privileged actions, the underlying API endpoints do not. Critical commands, such as starting/stopping the compressor, setting the pressure range, and resetting the device, can be invoked directly without credentials.
This design flaw completely bypasses all client-side security, allowing an unauthenticated attacker with network access to disrupt availability, manipulate pressure, or corrupt sensor data at will. 
Figure~\ref{fig:burpsuite} provides a stark example of this vulnerability, where a direct, unauthenticated POST request to the \texttt{/setparam} endpoint is accepted, allowing an attacker to modify critical safety thresholds at will.


\subsection{Denial Of Air Supply}

\begin{figure}[t!]
    \includegraphics[width=1\linewidth]{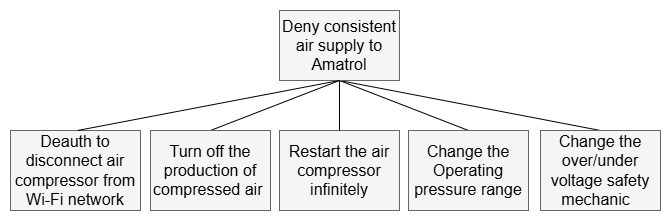}
    \caption{Impact $O_1$ Attack Tree: Denial of compressed air via unauthenticated API calls to power off the compressor, force continuous resets, set infeasible pressure thresholds, or misconfigure voltage safety limits.}
    \label{fig:impacto1}
\end{figure}

An attacker can disrupt normal network communication or exploit unauthenticated APIs to deny compressed air to the workcell, disrupting all pneumatic operations and achieving goal $O_1$ (Impact) as shown in Figure~\ref{fig:impacto1}.

\begin{figure}
    \centering
    \includegraphics[width=1\linewidth]{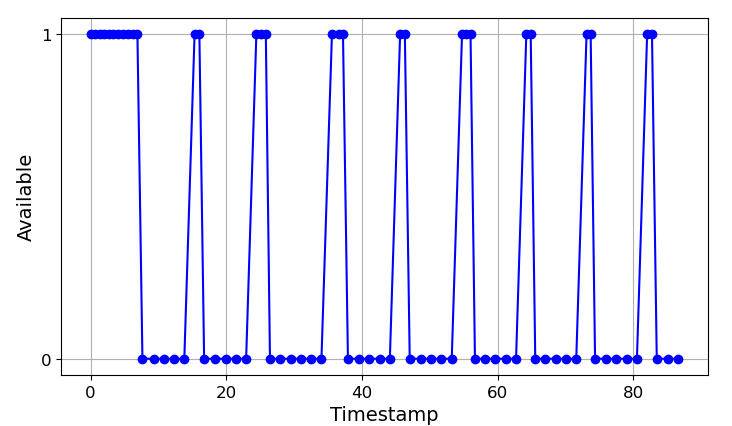}
    \caption{System availability during the Forced Reboot Loop attack ($AT_9$). The graph plots a binary availability metric (1 for available, 0 for unavailable) over time. After an initial stable period, the attacker begins sending repeated, unauthenticated \texttt{/reset} commands. Each drop to zero represents a successful device reboot, demonstrating a persistent denial-of-service condition where the compressor is unable to maintain operational uptime.}
    \label{fig:restartattack}
\end{figure}

\begin{itemize}
    \item ($AT_7$) \textbf{Wireless Availability Disruption}: Transmit spoofed deauthentication frames to force the controller offline, severing API connectivity and denying remote control and telemetry.
    \item ($AT_8$) \textbf{Unauthorized State Transition}: Continuously monitor the compressor state via the \texttt{/parameters} endpoint and invoke the \texttt{/off} command whenever the compressor becomes active.
    \item ($AT_9$) \textbf{Forced Reboot Loop}: Repeatably call the \texttt{/reset} endpoint to force constant reboots, preventing the compressor from ever reaching a stable, operational state. The resulting denial of service, which renders the device unavailable for extended periods, is quantified in Figure ~\ref{fig:restartattack}.
    \item ($AT_{10}$) \textbf{Setpoint Integrity Violation}: Use \texttt{/setpressurerange} to configure cut-in and cut-off thresholds to infeasible values (such as a cut-in pressure above the tank's maximum capacity), ensuring the compressor motor never engages.
    \item ($AT_{11}$) \textbf{Safety Threshold Manipulation}: Adjust the under- or over-voltage safety limits via the \texttt{/paramset} API to values that trigger an immediate protective shutdown.
\end{itemize}

\begin{figure}[!t]
    \centering
    \includegraphics[width=1\linewidth]{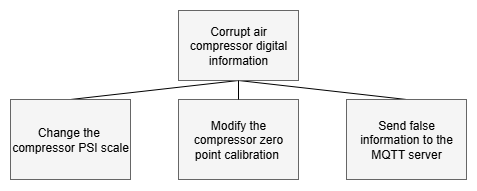}
    \caption{Impair Process Control $O_2$ Attack Tree: Manipulation of pressure telemetry via unauthenticated scale factor manipulation, zero-point offset tampering, and MQTT broker spoofing.}
    \label{fig:impairo2}
\end{figure}

\subsection{Corrupt Pressure Telemetry}

An attacker can undermine the integrity of the digital twin by injecting false pressure readings, achieving goal $O_2$ (Impair Process Control), as shown in Figure~\ref{fig:impairo2}. These attacks cause operators and automated systems to misinterpret the system's state. 
Related attacks include:


\begin{figure}[t!]
    \centering
    \includegraphics[width=1\linewidth]{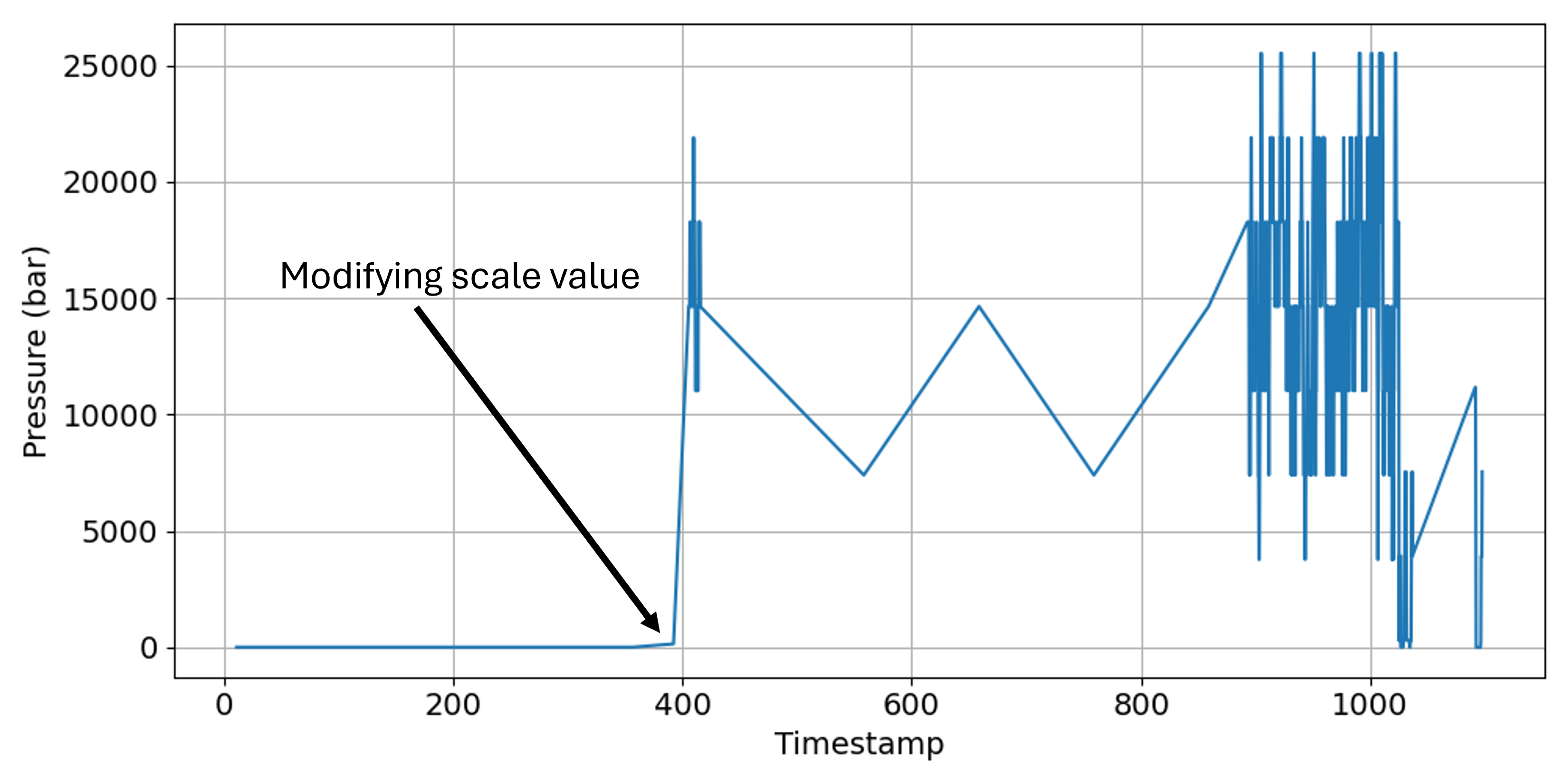}
    \caption{ Impact of a Sensor Calibration Manipulation attack ($AT_{12}$). The reported pressure telemetry exhibits erratic and unstable fluctuations after an attacker applies a malicious scaling factor via the \texttt{/calibrate} API. This manipulation effectively conceals the true, stable pressure within the system, deceiving monitoring tools and potentially masking legitimate physical-layer events.}
    \label{fig:calibrationattackgraph}
\end{figure}

\begin{figure}[!t]
    \centering
    \includegraphics[width=1\linewidth]{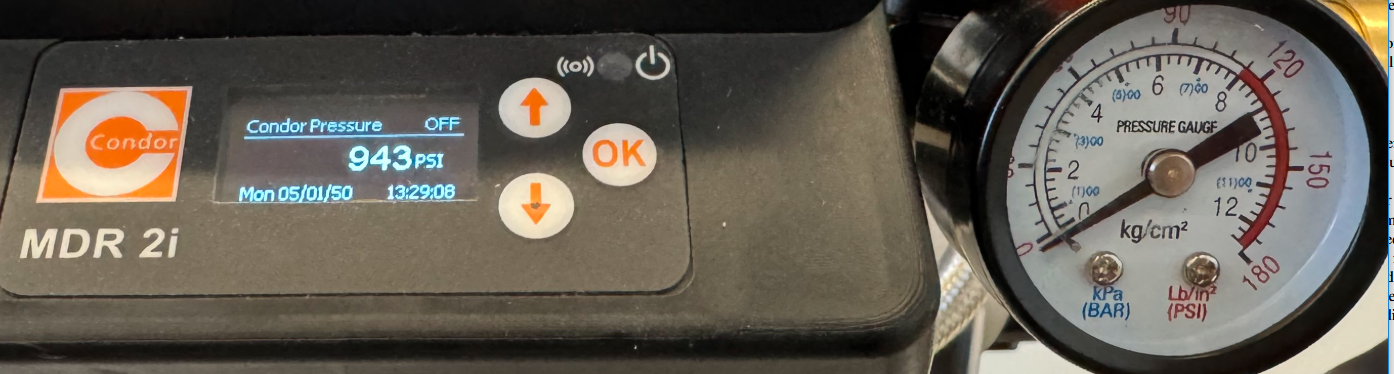}
    \caption{Impact of a Sensor Zero-Point Manipulation attack ($AT_{13}$). While the physical analog gauge correctly reads 0 PSI, the attacker has shifted the sensor's baseline, causing the digital display to report a dangerously inaccurate 943 PSI.}
    \label{fig:comparereaders}
\end{figure}

\begin{itemize}
    \item ($AT_{12}$) \textbf{Sensor Calibration Manipulation}: The \texttt{/calibrate} API applies a scaling factor to the raw pressure sensor output. By setting an arbitrary scale, an attacker can conceal the true tank pressure, for example, by making it report double or half the actual value. The resulting erratic and deceptive telemetry is shown in Figure~\ref{fig:calibrationattackgraph}.
    \item ($AT_{13}$) \textbf{Sensor Zero-Point Manipulation}: The \texttt{/calibratezeropoint} API shifts the sensor's baseline. An attacker can introduce an offset that causes the sensor to report a non-zero pressure when the tank is empty or an artificially zero value when it is full, invalidating all subsequent readings. Figure~\ref{fig:comparereaders} provides a stark example of this attack, where the controller falsely reports 943 PSI while the actual tank pressure is zero.
    \item ($AT_{14}$) \textbf{External Telemetry Injection}: In configurations where the compressor publishes telemetry to an MQTT broker, an attacker on the same network can inject fabricated pressure messages, poisoning the data stream for all subscribers.
\end{itemize}

\section{Defensive Measures and Mitigations}
\label{sec:defense}

To mitigate the attacks demonstrated in this paper, we propose a series of layered defenses. 
Each defense is designed to counter one or more of the attack vectors identified in Section~\ref{sec:evaluation}.
A comprehensive traceability matrix mapping the threat model, attacks, and these defenses is provided in Table~\ref{tab:traceability}.
The following subsections detail these defensive measures, which are mapped to the MITRE Common Weakness Evaluation (CWE~\cite{CWE}) and the Common Attack Pattern Enumeration and Classification (CAPEC~\cite{CAPEC}) frameworks.

\begin{table*}[t!]
\centering
\caption{Traceability Matrix: From Threat Model to Attacks and Mitigations}
\label{tab:traceability}
\begin{tabularx}{\textwidth}{cXX}
\toprule
\textbf{Trust Boundary (TB)} & \textbf{Vulnerabilities \& Attacks (AT)} & \textbf{Primary Defenses (D)} \\
\midrule
\textbf{$TB_1$: Wireless Interface} & 
    \textbullet{} \textbf{$AT_1$}: Hardcoded Network Credentials \newline
    \textbullet{} \textbf{$AT_2$}: Station Mode Reconnaissance \newline
    \textbullet{} \textbf{$AT_7$}: Wireless Availability Disruption
& 
    \textbullet{} \textbf{$D_1$}: Unique Per-Device Credentials \newline
    \textbullet{} \textbf{$D_5$}: Management Frame Protection \newline
    \textbullet{} \textbf{$D_6$}: WPA3 Network Security \newline
    \textbullet{} \textbf{$D_9$}: Authentication Event Auditing
\\ 
\addlinespace
\textbf{$TB_2$: Web App \& API} & 
    \textbullet{} \textbf{$AT_3$}: Unencrypted Control Plane \newline
    \textbullet{} \textbf{$AT_4$}: Unrestricted Credential Brute-Force \newline
    \textbullet{} \textbf{$AT_5$}: Use of Shared Static Credentials \newline
    \textbullet{} \textbf{$AT_6$}: Missing API Endpoint Authentication \newline
    \textbullet{} \textbf{$AT_8$-$AT_{14}$}: All API-based attacks
& 
    \textbullet{} \textbf{$D_1$-$D_4$}: Credential Lifecycle Management \newline
    \textbullet{} \textbf{$D_7$}: Transport Layer Security (TLS) \newline
    \textbullet{} \textbf{$D_8$}: Brute-Force Protection \newline
    \textbullet{} \textbf{$D_9$}: Authentication Event Auditing \newline
    \textbullet{} \textbf{$D_{10}, D_{11}$}: API Authentication \& Authorization \newline
    \textbullet{} \textbf{$D_{12}$}: Separate Control and Management Planes
\\ 
\addlinespace
\textbf{$TB_3$: Controller to Plant} & 
    \textbf{Indirectly Violated}: This boundary is compromised by attacks crossing $TB_2$ (e.g., $AT_{10}, AT_{11}$) that send malicious commands to the electro-pneumatic domain.
& 
    \textbullet{} Defenses for $TB_2$ (\textbf{$D_{10}, D_{11}$}) prevent unauthorized commands. \newline
    \textbullet{} \textbf{$D_{12}$} provides the primary architectural mitigation by removing the dangerous API surface entirely.
\\ 
\addlinespace
\textbf{$TB_4$: OTA Update Path} & 
    \textbf{Inferred Vulnerability}: Lack of secure boot, cryptographic signature verification, or a hardware root of trust. \textit{(This vector was not exploited in this work.)}
& 
    \textbullet{} \textbf{$D_{13}$}: Secure Firmware Update Process \newline
    \textbullet{} \textbf{$D_{14}$}: Hardware Root of Trust \\ 
\bottomrule
\end{tabularx}
\end{table*}

\subsection{Harden Device Credentials} 
The use of hardcoded and shared credentials (CWE-798: Use of Hard-coded Credentials) enables trivial privileged access. To mitigate this, manufacturers should:
\begin{itemize}
    \item ($D_1$) \textbf{Unique Per-Device Credentials}: Each device must ship with a unique, randomly generated password for both its Wi-Fi AP mode and its administrative roles.
    \item ($D_2$) \textbf{Mandatory First-Use Credential Rotation}: Require the user to change the default password upon initial setup.
    \item ($D_3$) \textbf{Enforce Credential Complexity Policies}: Enforce complexity requirements and periodic rotation for all privileged accounts to mitigate weak password choices (CWE-521).
    \item ($D_4$) \textbf{Removal of Hardcoded Backdoors}: Ensure no hidden backdoor or maintenance accounts exist that are not documented for the end-user (CWE-912: Hidden Functionality). All accounts providing privileged access must be under the asset owner's control to ensure accountability and facilitate security audits.
\end{itemize}

\subsection{Secure Network Communications} 
The device's network interfaces are vulnerable to eavesdropping and disruption.
\begin{itemize}
    \item ($D_5$) \textbf{Implement Management Frame Protection}: Wireless deauthentication is a well-known denial-of-service vector. Implementing Management Frame Protection (MFP), as specified in IEEE 802.11w, is essential to protect the availability of the wireless connection.
    \item ($D_6$) \textbf{Adopt WPA3 Network Security}: Adding WPA3 provides stronger protection against offline password cracking and ensures forward secrecy for network traffic.
    \item ($D_7$) \textbf{Enforce Transport Layer Security (TLS)}: All communication with the web interface must be protected with HTTPS (TLS). This mitigates the passive sniffing of credentials and control commands (CWE-319: Cleartext Transmission of Sensitive Information) and thwarts network eavesdropping attacks (CAPEC-94).
\end{itemize}

\subsection{Strengthen Role-Based Authentication}
The lack of protection against brute-force attacks (CAPEC-112) is a critical flaw. This can be addressed by:
\begin{itemize}
    \item ($D_8$) \textbf{Implement Brute-Force Protection}: Introduce account lockout policies after a set number of failed login attempts to frustrate automated brute-forcing (CWE-307: Improper Restriction of Excessive Authentication Attempts).
    \item ($D_9$) \textbf{Implement Authentication Event Auditing}: All successful and failed authentication attempts should be logged. This provides visibility into potential attacks and supports forensic investigation.
\end{itemize}

\subsection{Require Authentication for all API Calls}
The presence of unauthenticated API endpoints (CWE-306: Missing Authentication for Critical Function) is the root cause of the most severe impact scenarios. 
\begin{itemize}
    \item ($D_{10}$) \textbf{Require Per-Request API Authentication}: Every API endpoint that performs a state-changing or sensitive operation must require a session token or other authentication mechanism that is validated on the server side.
    \item ($D_{11}$) \textbf{Enforce Server-Side Authorization}: API authorization should be enforced on the server, ensuring that users can only perform actions permitted by their role.
    \item  ($D_{12}$) \textbf{Architecturally Separate Control and Management Planes}: The device insecurely comingles its routine control plane (start/stop, telemetry) with its highly privileged management plane (sensor calibration, safety threshold modification) on a single, unauthenticated API. This constitutes an instance of CWE-749 (Exposed Dangerous Method or Function). A secure design must architecturally separate these planes. Process critical functions such as operating pressure ranges (\texttt{/setpressurerange}), safety thresholds (through \texttt{/paramset}), and sensor calibration (\texttt{/calibrate}) must be restricted to a dedicated, physically-secured management interface (e.g., a local serial or USB port) that is disabled during normal operation. The network API should be strictly limited to essential operating functions, enforcing a clear separation of duties between operators and maintainers.   
\end{itemize}

\subsection{Ensure Secure Firmware Updates}
The OTA update mechanism presents a critical trust boundary (TB4) that, if compromised, could allow an attacker to gain persistent control.
\begin{itemize}
    \item ($D_{13}$) \textbf{Implement Secure Firmware Updates}: The controller must cryptographically verify the signature of any new firmware image before applying it. This ensures the update is from a trusted source (authenticity) and has not been tampered with (integrity), mitigating attacks like those described in CAPEC-440 (Upload of Malicious File).
    \item ($D_{14}$) \textbf{Implement Hardware Root of Trust}: To anchor the security of the boot and update process, the design should incorporate a Hardware Root of Trust (HRoT), such as a Trusted Platform Module (TPM) or a secure element. An HRoT provides a protected environment for cryptographic operations, including the secure storage of cryptographic keys used for signature verification. This ensures that the firmware validation process itself can be trusted, even if the main processor's software environment is compromised, and it provides a strong foundation for a secure boot process.
\end{itemize}

\section{Supply Chain Case Study}
\label{sec:supplychain}

The security vulnerabilities detailed in Section~\ref{sec:evaluation}, from the hardcoded credentials ($AT_1$, $AT_5$) to the complete lack of API authentication ($AT_6$), are not isolated technical defects. They are direct artifacts of the fragmented, multi-party supply chain responsible for the product's design and integration. The presence of fundamental weaknesses is symptomatic of the systemic gaps in security requirement definition, accountability, and validation processes that we analyze in this section.

\subsection{Responsible Disclosure}

All vulnerabilities described in this paper were reported following coordinated vulnerability disclosure (CVD) best practices~\cite{CVD2017}. Beginning 120 days before publication, we attempted direct disclosure to California Air Tools, but calls were disconnected without response and emails were never received a proper repsponse. Condor-USA redirected us back to Condor-Werke GmbH, from whom we received no acknowledgment. To ensure end-user safety and meet disclosure timelines, we engaged CISA through Carnegie Mellon University’s SEI CVE program via the VINCE coordinated disclosure platform~\cite{VINCE}. 

\subsection{Supply Chain Structure Analysis}

Our investigation revealed a complex supply chain involving three distinct entities with complementary but non-overlapping areas of expertise:

\begin{itemize}
\item \textbf{California Air Tools:} Equipment integrator specializing in pneumatic tools and compressors, with established expertise in mechanical engineering and manufacturing processes.
\item \textbf{Condor-USA:} U.S. distributor for German-manufactured pressure control components, serving as the commercial interface for North American markets.
\item \textbf{Condor-Werke GmbH:} German OEM manufacturer with decades of experience in mechanical pressure control systems, expanding into IoT-enabled products for the first time with this product line.
\end{itemize}

This structure represents a common pattern in industrial equipment manufacturing, where traditional mechanical expertise is combined with emerging connectivity capabilities through supply chain partnerships. However, this distribution of expertise can create challenges in security coordination.

\subsection{Observed Security Process Gaps}

Our coordinated disclosure process revealed systematic gaps that directly explain the origin of the vulnerabilities found in Section~\ref{sec:evaluation}. For example, the lack of API authentication ($AT_6$) can be traced to deficiencies in requirements specification, while the continued existence of hardcoded PINs ($AT_5$) is a result of fragmented responsibility and a lack of security validation. The following gaps were observed:

($G_1$) \textbf{Requirements Specification Deficiencies}: The procurement specifications lacked explicit cybersecurity requirements, performance criteria, or validation procedures for the IoT components. This pattern reflects broader industry challenges where traditional manufacturers must adapt procurement processes for connected products without established cybersecurity frameworks.

($G_2$) \textbf{Expertise Integration Challenges}: The addition of IoT capabilities to established mechanical products creates integration challenges when cybersecurity expertise is distributed across multiple organizations. In our case, the Condor-Werke MDR2i controller appears to be one of their first IoT-enabled offerings, demonstrating the learning curve that specialized manufacturers face when transitioning legacy product lines to connected solutions.

($G_3$) \textbf{Responsibility Attribution Complexity}: The difficulty in assigning ownership for remediation was a primary reason that fundamental vulnerabilities such as hardcoded credentials ($AT_1$, $AT_5$) persisted. Each organization appropriately focused on its core domain; specifically mechanical design, pressure control, or system integration. However, no single party assumed end-to-end security ownership. This fragmentation delayed coordinated response and illustrates the need for clearly defined security roles in multi-party IoT deployments.

($G_4$) \textbf{Communication Channel Limitations}: The multi-tiered structure created challenges in establishing direct communication channels for security issues. Initial outreach was routed through standard commercial contacts rather than a dedicated vulnerability response channel, resulting in dropped or redirected inquiries. Establishing formal incident response protocols that span all supply chain tiers is essential to ensure timely and effective security coordination.

($G_5$) \textbf{Market Timing Pressures.} Competitive pressure to add “smart” capabilities to traditional products led to abbreviated security validation processes, compounding integration and accountability challenges.

\subsection{Regulatory and Legal Considerations}

The international nature of this supply chain introduces regulatory complexity that can contribute to security gaps. 
Condor-Werke's operations in Germany subject them to EU cybersecurity frameworks, while California Air Tools operates under US regulations. 
The EU Cyber Resilience Act, which entered into force on December 10, 2024, with full enforcement beginning December 11, 2027, would likely require many of the security controls recommended in Section VI for future products~\cite{EU2024CRA}. 
However, products developed before such regulations take effect may remain vulnerable without the proactive intervention of the manufacturer.

\subsection{Recommendations for Supply Chain Security}

Based on this case study analysis, we recommend the following systematic improvements to supply chain security practices:

\textbf{Security Requirements Integration.} To counter the lack of upfront security validation ($G_1$), system integrators should develop standardized cybersecurity requirements for connected device procurement, including specific technical controls, validation procedures, and ongoing support commitments. These requirements should be proportional to the operational criticality of the target application~\cite{NIST2022SCRM}.

\textbf{Capability Assessment and Development.} To address the challenge of integrating disparate skill sets across organizations ($G_2$), organizations engaging in IoT device development should conduct systematic cybersecurity capability assessments across their supply chain, identifying gaps and establishing development or partnership strategies to address them. This may include third-party security validation for organizations new to connected product development.

\textbf{Shared Responsibility Frameworks.} To solve the problem of fragmented ownership and accountability ($G_3$), clear contractual frameworks should establish security responsibilities across supply chain relationships, including incident response procedures, vulnerability management processes, and long-term update support commitments. These frameworks should account for the distributed nature of expertise in complex supply chains.

\textbf{Security-Focused Communication Channels.} To overcome the communication silos that hindered the disclosure process ($G_4$), supply chains should establish dedicated security communication channels that enable direct coordination between technical security teams across organizational boundaries, supplementing commercial relationship channels.

\textbf{Lifecycle Security Planning.} To mitigate risks from abbreviated development cycles driven by market pressures ($G_5$), security planning should extend beyond initial product development to include long-term support, update mechanisms, and end-of-life security considerations throughout the anticipated product lifecycle.



\section{Related Work}
\label{sec:relatedwork}
Cyber-Physical Systems (CPS) are a new generation of systems with integrated computational and physical capabilities~\cite{baheti2011cyber}. Their adoption has accelerated as factories strive to meet Industry 4.0 standards, but this has also introduced new attack surfaces and made industrial environments more vulnerable to cyber attacks~\cite{alguliyev2018cyber}. This section discusses related work in IIoT security, attack trends, and defensive frameworks.

\subsection{IIoT Security}
\label{sec:relatedworkIIoTSecurity}

Industrial Internet of Things are becoming increasingly prevalent in factories worldwide, raising significant security concerns. Studies highlight that IIoT devices often suffer from insecure firmware, weak authentication, and limited update mechanisms, making them attractive targets for attackers~\cite{fabri2023industrial}. 
Our analysis of the CAT-10020SMHAD smart air compressor provides a concrete example of these exact issues in a commercial product. The use of a hardcoded, unchangeable PIN and a default, publicly documented Wi-Fi password is a textbook case of weak authentication. Likewise, the controller's vulnerability to attacks on its OTA update process (Trust Boundary $TB_4$) exemplifies the risks of limited update mechanisms.

Although these general weaknesses are well documented in surveys, detailed case studies of commercial-off-the-shelf industrial equipment demonstrating the practical impact of these flaws are less common. This work helps fill that gap by providing a practical demonstration of how these specific vulnerabilities, in addition to unauthenticated APIs and unencrypted communications, translate into the operational risks of process disruption ($O_1$) and control impairment ($O_2$) in an industrial component.

\subsection{Attack Trends in Industrial Settings} 
\label{sec:relatedworkAttackTrends}

Attacks targeting CPS and IIoT in industrial contexts have become increasingly sophisticated. 
The Stuxnet worm famously demonstrated how malicious code could manipulate PLCs to cause physical damage~\cite{baezner2017stuxnet}, while the more recent FrostyGoop malware was the first reported instance of Modbus/TCP abuse in the wild to achieve physical impact~\cite{parsons2024frostygoop}. 
As highlighted by Panchal et al., attack surfaces in OT are often layered, spanning from low-level sensors and embedded devices to SCADA and HMI systems~\cite{Panchal2018}. 
Our work provides a practical demonstration of threats at these lower layers. 
The attacks against the smart air compressor align directly with established vectors in the literature:
\begin{itemize}
    \item The manipulation of calibration and zero-point offsets is a form of \textit{sensor spoofing}, similar in principle to the Hall Spoofing attack described by Barua et al., where sensor output is manipulated to deceive control logic~\cite{255312}.
    \item The use of unauthenticated API calls to remotely start, stop, or reset the compressor is a form of \textit{malicious command injection}, akin to how Industroyer sent commands to circuit breakers~\cite{slowik2019crashoverride}.
    \item By corrupting telemetry data, an attacker can perform a \textit{stealthy deception attack}, misleading a human operator or digital twin about the process state, similar to the HMI communication hijacking shown by Kleinmann et al. \cite{kleinmann2017stealthy}.
\end{itemize}
This case study therefore provides a tangible example of how these theoretical attack patterns manifest in a commercial IIoT device, bridging the gap between general surveys and practical, real-world impact.

\subsection{Industrial Security Frameworks}

To counter industrial cyber threats, several industry- and government-led frameworks provide guidance. The ISA/IEC 62443 series of standards offers a comprehensive framework for securing Industrial Automation and Control Systems (IACS), defining requirements for components, systems, and processes~\cite{IEC62443-4-2-2019, ISCI2024}. This paper's threat model explicitly adopts the Security Level (SL) concept from this standard to define the threat actor. Similarly, the NIST Cybersecurity Framework (CSF) and its operational technology specialization in NIST SP 800-82 Rev. 3 provide a risk-based approach to secure OT environments~\cite{NIST-SP800-82r3}. The defensive recommendations in this paper are directly aligned with the technical controls and principles outlined in these foundational documents, grounding our findings in established best practices.



\section{Conclusion}
\label{sec:conclusion}

As companies rush to modernize their factory lines with IIoT devices in pursuit of Industry 4.0 standards, they risk exposing themselves to new and significant attack surfaces. This paper presented a case study in which the pursuit of market trends led to the development of an IIoT-based cyber-physical system where fundamental security principles were overlooked. These risks were amplified by a fragmented supply chain where multiple organizations, each new to secure software development, collaborated without a clear framework for end-to-end security ownership. 

Our experimental evaluation demonstrated that fundamental security deficiencies can be exploited by an adversary with SL-1 capabilities to cause significant operational impacts, including process disruption and the corruption of critical telemetry. More broadly, this case study is emblematic of a systematic challenge in industrial manufacturing where the security posture of a final product is bounded by the least mature security practices within its supply chain. The findings underscore that technical security controls at the component level are insufficient and must be augmented with robust supply chain governance to achieve resilient industrial systems. Ensuring the promise of Industry 4.0 is contingent upon a commitment to security-by-design, transparent communication, and share responsibility among all ecosystem stakeholders.

\bibliographystyle{IEEEtran}
\bibliography{IEEEabrv,refs}

\end{document}